\documentstyle[12pt,aaspp4]{article}
\makeatletter
\@ifundefined{chapter}{\def\thebibliography#1{\section*{References\@mkboth
  {REFERENCES}{REFERENCES}}\list
  {\relax}{\setlength{\labelsep}{0em}
        \setlength{\itemindent}{-\bibhang}
        \setlength{\itemsep}{0pt}
        \setlength{\parsep}{0pt}
        \setlength{\leftmargin}{\bibhang}}
    \def\newblock{\hskip .11em plus .33em minus .07em}
    \sloppy\clubpenalty4000\widowpenalty4000
    \sfcode`\.=1000\relax}}%
{\def\thebibliography#1{\chapter*{Bibliography\@mkboth
  {BIBLIOGRAPHY}{BIBLIOGRAPHY}}\list
  {\relax}{\setlength{\labelsep}{0em}
        \setlength{\itemindent}{-\bibhang}
        \setlength{\itemsep}{0pt}
        \setlength{\parsep}{0pt}
        \setlength{\leftmargin}{\bibhang}}
    \def\newblock{\hskip .11em plus .33em minus .07em}
    \sloppy\clubpenalty4000\widowpenalty4000
    \sfcode`\.=1000\relax}}

\newlength{\bibhang}
\let\@internalcite\cite
\def\cite{\let\@citeleft(\let\@citeright)%
    \@ifstar{\citeyear}{\citefull}}
\def\acite{\let\@citeleft\relax\let\@citeright\relax%
    \@ifstar{\citeyear}{\acitefull}}
\def\citenp{\let\@citeleft\relax\let\@citeright\relax
    \@ifstar{\citeyear}{\citefull}}
\def\citefull{\def\astroncite##1##2{##1~##2}\@internalcite}
\def\citeyear{\def\astroncite##1##2{##2}\@internalcite}
\def\acitefull{\def\astroncite##1##2{##1~(##2)}\@internalcite}

\def\@citex[#1]#2{\if@filesw\immediate\write\@auxout{\string\citation{#2}}\fi
  \def\@citea{}\@cite{\@for\@citeb:=#2\do
    {\@citea\def\@citea{; }\@ifundefined
       {b@\@citeb}{{\bf ?}\@warning
       {Citation `\@citeb' on page \thepage \space undefined}}%
{\csname b@\@citeb\endcsname}}}{#1}}

\def\@cite#1#2{\@citeleft#1\if@tempswa , #2\fi\@citeright}
\def\@biblabel#1{}

\makeatother

\newcommand{\PSbox}[3]{\mbox{\rule{0in}{#3}\includegraphics{#1}\hspace{#2}}}
\newcommand{\FigNum}[1]{\unitlength 1pt \begin{picture}(55,10)(-400,35) 
                        \put(0,0){Figure #1}
                        \end{picture}}

\newcommand{\persec}{\mbox{$\second^{-1}$}}
\newcommand{\percm}{\mbox{$\cm^{-2}$}}
\newcommand{\ppm}{\mbox{$\pm$}}
\newcommand{\cgsflux}{\erg\percm\persec}

\newcommand\approxlt{\mbox{$^{<}\hspace{-0.24cm}_{\sim}$}}

\newcommand{\ee}[1]{\mbox{$10^{#1}$}}
\newcommand{\tee}[1]{\mbox{$\times 10^{#1}$}}

\newcommand{\perval}[2]{{#1\mbox{$^{#2}$}}}

\def\x1608{{4U~1608$-$522}}

\def\cenx4{{Cen~X$-$4}}

\def\saxj1808{{SAX J1808.4$-$3658}}
\newcommand{\cm}{\mbox{$\rm\,cm$}}

\newcommand{\second}{\mbox{$\rm\,s$}}

\newcommand{\erg}{\mbox{$\rm\,erg$}}

\newcommand{\chandra}{{\em Chandra\/}}

\newcommand{\rosat}{{\em ROSAT\/}}

\newcommand{\xmm}{{\em XMM\/}-Newton}
\newcommand{\beppo}{{\em BeppoSAX\/}}
\newcommand{\sax}{\beppo}

\newcommand{\hhh}{{H02}}

\newcommand{\cgsfluence}{\mbox{${\rm erg}\, {\rm cm^{-2}}\, {\rm event^{-1}}$}}
\newcommand{\totalevents}{events$\,$\perval{sky}{-1}$\,$\perval{yr}{-1}}

\begin{document}

\title{A Search for X-Ray Flashes with \xmm\ }

\author{Nicholas M. Law\altaffilmark{1,2}, Robert E. Rutledge\altaffilmark{2} and Shrinivas R. Kulkarni\altaffilmark{2}, 
}
\altaffiltext{1}{
Institute of Astronomy, Madingley Road, Cambridge CB3 0HA, UK; nml27@ast.cam.ac.uk
}
\altaffiltext{2}{
Division of Physics, Mathematics and Astronomy, 
California Institute of Technology, MS 130-33, Pasadena, CA 91125;
rutledge@tapir.caltech.edu, srk@astro.caltech.edu 
}

%\maketitle

\begin{abstract}

We searched for X-ray flashes (XRFs) -- which we defined as
$\sim$10~s duration transient X-ray events observable in the 0.4-15
keV passband -- in fields observed using \xmm\ with the EPIC/pn
detector.  While we find two non-Poissonian events, the astrophysical
nature of the events is not confirmed in fully simultaneous
observations with the EPIC/MOS detectors, and we conclude that the
events are anomalous to the EPIC/pn detector.  We find a 90\% upper
limit on the number of flashes per sky per year at two different
incoming flash fluxes: 4.0\tee{9}~\totalevents\ for a flux of
7.1\tee{-13} \cgsflux\ and 6.8\tee{7}~\totalevents\ for 1.4\tee{-11}
\cgsflux, both assuming a spectral power-law photon index $\alpha$=2.
These limits are consistent with an extrapolation from the \sax/WFC
XRF rate at much higher fluxes ($\sim$\,a factor of \ee{5}), assuming an
homogenous population, and with a previous, more stringent limit
derived from ROSAT pointed observations.
\vspace{0.5cm}
\\
This is a preprint of an Article accepted for publication in
{\it Monthly Notices of the Royal Astronomical Society} \copyright\, 2004 The Royal Astronomical Society
\end{abstract}

\section{Introduction}

Observations of large areas of the sky in the X-ray (1-20 keV) and
gamma-ray (20-1000 keV) passbands have found bursts of X-rays -- X-ray
flashes (XRFs). The properties of XRFs are reviewed in (for example)
\acite{heise01c} and are observationally distinct from (but may be
related to) the ubiquitous and relatively well-studied gamma-ray
bursts (GRBs).  The phenomenological definition of these events,
proposed by \acite{heise02} (\hhh\ hereafter), includes strong 2-10
keV emission, in which the 2-10 keV fluence is greater than the 50-300
keV fluence; a duration of \approxlt\  few \ee{3}~s (to distinguish from
flare-stars); and the absence of a strong optical or IR counterpart
(to distinguish from X-ray binaries and coronally active stars).
These phenomena have recently been surveyed \hhh\ using \sax/WFC
observations (2-25 keV sensitivity) over 6 years and a 40$\times$40 sq
deg field of view.  This resulted in a detection of 34 of these
events, with typical X-ray fluxes of \ee{-8} \cgsflux\ (2-20 keV). 
\citenp{heise01} and \citenp{heise01b} describe XRFs in detail.

XRFs can be observationally distinguished from the well-studied type-I
X-ray bursts, which are due to thermonuclear flashes on the surfaces
of neturon stars (NSs) in accreting low-mass X-ray binaires (see
\citenp{lvt93} for a review); the type-I X-ray bursts have
$\approx1\,{\rm s}$ rise-times and exponential decays with
characteristic timescales of 10-$100\,{\rm s}$, during which the thermal
spectrum softens as the NS atmosphere cools.  XRFs, by contrast, have
non-thermal spectra, and their lightcurves typically do not follow a
fast-rise/exponential decay time profile.

At present, the origin of the XRFs is unclear. Some GRB models permit
similar bursts at lower photon energies, by altering just one
parameter of the model.  An example of such a model is the so-called
hypernova model, in which the details of the characteristics of the
progenitor star (e.g. mass, spin) may play a critical
role in determining the energy band of the prompt emission.  In such a
case, the XRFs would originate from a similar parent population to the
GRBs (at cosmological distances, in star-forming galaxies), and thus
share some observational properties with GRBs -- such as an homogenous
distribution on the sky, and a break in the logN--logS cumulative
distribution due to cosmological (and, perhaps, source population)
evolution.  Another example of a GRB-origin model which can produce
the XRFs is to place the GRBs at high redshift, so that the spectral
energy distribution peaks at a lower energy.  However, the optical
detection of the host galaxies associated with XRF~011030
\cite{fruchter02a}
 and XRF 020427 \cite{fruchter02b} 
make this explanation less likely.  Other GRB models -- such as
inspiralling binary NSs, which may not vary in the energy band where
most of their energy is emitted -- may not be able to accommodate the
XRF phenomena, in which case the XRFs could originate from a distinct
population.

Data accumulated from X-ray satellites may contain previously
unrecognized X-ray flashes (unresolved transients lasting O{[}10-100
s{]}).  Events with characteristics similar to X-ray flashes have been
claimed detected in {\em Einstein} observatory data \cite{gotthelf96},
down to \ee{-11} \cgsflux\ in 1.5\tee{7}$\,{\rm s}$ of data (0.2-3.5
keV) for a $1\,\deg^2$ FOV, for a total integration of 1.5\tee{7}$\,{\rm
s}\,\deg^{2}$, in which 42 events were detected; the implied burst
rate is 3.7\tee{6} \totalevents\ at a fluence of 2\tee{-10}
\cgsfluence.  A search with \rosat/PSPC \cite{vikhlinin98} with a
comparable exposure and flux limit (1.6\tee{7}$\,{\rm s}$,
2.7$\,\deg^2$ FOV, 0.1--2.4 keV) found no bursts, producing a 90\%
confidence upper-limit ($<$2 bursts) of 6.1\tee{4} \totalevents\ above
a fluence of 2.6\tee{-10} \cgsfluence, which contradicts the {\em
Einstein} result.

The new generation of sensitive X-ray observatories (\chandra, \xmm)
offer a new opportunity to search for these events with detectors having
new characteristics, most notably in spectral response. In particular, the intrinsic absorption of X-ray flashes is unknown;
if the absorption in X-ray flashes were, on average, high ($10^{22}
{\rm cm}^{-2}$) then a low detected rate with ROSAT (0.1-2 keV) could
still translate in to a detected XRF rate in excess of that expected
from a an $N(>F)\propto F^{-3/2}$ law.

We performed such
a search in a number of observations from \xmm-EPIC/pn; 
although we find two non-Poissonian
events, their astrophysical origin is not confirmed with fully
simultaneous observation with \xmm-EPIC/MOS; we conclude that the
events are anomolous to the EPIC/pn detector.  We subsequently set
upper-limits for the full-sky XRF rate.

The paper is organized as follows.  In \S~\ref{sec:feasibility}, we
estimate detectable burst rates for sensitive X-ray observatories,
finding that the most sensitive instrument to the phenomenon is the
XMM-Newton EPIC/pn detector.  In \S~3, we describe the \xmm\ dataset
we used, and the detection algorithm, together with source
characterization procedures and XRF detection sensitivity calculations. 
In \S~4 we give the results of our search, and describe the XRF detection 
sensitivity.  We discuss these results in \S~5 and conclude in \S~6.

\section{Observational Burst Rate Estimates}
\label{sec:feasibility}

The feasibility of detection of XRFs with a particular instrument may
be extrapolated from the number of flashes detected with SAX/WFC
(\hhh).  Assuming the objects producing the flashes are isotropically
and homogenously distributed the number of detectable flashes scales
as

{\centering \begin{equation}
\label{eq:flashrateextrapolation}
\frac{N(>F_{1})}{N(>F_{2})}={\left( \frac{F_{1}}{F_{2}}\right) }^{-\frac{3}{2}}
\end{equation}
\par}

where $F_{1}$ and $F_{2}$ are the flux limits of the detectors and
$N(>F_{1})$ and $N(>F_{2})$ are the numbers of flashes detected at
those flux limits.

In this feasibility study we assume a
significant detection requires a minimum flux corresponding to one
count per second at the detector chip at the peak of the XRF.  This flux
was found from
WebPIMMS\footnote{http://heasarc.gsfc.nasa.gov/Tools/w3pimms.html} and
is dependent on the spectral shape of the flash, which we took to be a
photon spectral slope between $\alpha$=1.2--2.0, as observed with
SAX/WFC \cite{heise01}. Data from the Chandra ACIS and the XMM-Newton
EPIC/pn and MOS cameras are available to us in public archives. Using
Eqn.~\ref{eq:flashrateextrapolation} the total XRFs per sky per year
at a given flux rate (and so a given detector sensitivity) were
extrapolated from those observed with SAX/WFC.  Using the power-law
photon index range observed with SAX/WFC, the solid angle
field-of-view of the detectors, and the total integrated time using
each detector, we calculated the expected detectable number of events
(Table \ref{tab: feasibility}) in the presently available data.

The XMM-Newton EPIC/pn detector offers the greatest chance of flash
detection in available data. We focus on the EPIC/pn detector, but it
should be noted that our algorithms are applicable to other datasets
and can be used, for example, without modification with EPIC MOS data.
Depending on the spectrum, the EPIC/MOS1+2 detectors (together) have
$\sim$50\% of the effective area of the EPIC/pn detector. As the MOS
and pn detectors observe the same region of sky simultaneously,
coincidence of events on both detectors may be used for confirmation
of the astrophysical origin of transients.

\section{XMM-Newton Data and Analysis}

\subsection{\label{sec: data set}Data set}

Observations were selected from the \xmm\ Science Data
Archive\footnote{\xmm\ Science Operations Center,
http://xmm.vilspa.esa.es} based on the following criteria:

\begin{enumerate}
\item We use observations publicly or otherwise available to us on 2002 June 1; 
\item Observations with $>$ 50 ks total observation time (including
periods in which the detector was switched off). Extension to
smaller observation times increases the chance of flash detection
proportional to the total integration time, but at an increased
overhead of file handling and data collation; 50 ks was chosen as
a compromise value. 
\item Observations within 15$^\circ$ latitude to the Galactic
equator were excluded to reduce the incidence of transients from Galactic
sources (X-ray binaries, for example). 
\end{enumerate}

Applying these criteria we analysed 225.1\,ks of observations, split into 6
separate observations with total detector-on times ranging from 19939\,s to
47721\,s. The algorithms described below can be applied to new data as
it becomes available.

Simulations using WebPIMMS and the observed background spectrum
suggested that detection sensitivity would be increased by removing
counts with energies below 0.4~keV.  We therefore only consider counts
between 0.4 and 15 keV. 

\subsection{XRF Detection}

An outline of our XRF detection algorithm is as follows. We split
observation event lists into short sections of time, with duration of
500$\,{\rm s}$, with 100$\,{\rm s}$ of overlap between adjacent
sections.  Within each time section, we bin counts into a $256\times
256\times256$ bin data cube. Each bin volume is
9.95\arcsec$\times$9.95\arcsec\ spatially and has a duration of
$2.34\,{\rm s}$. We thus oversample both the point-spread-function
(PSF; $\sim$50\arcsec) and the desired optimal burst duration. We took
this duration to be 10\,s, which importantly affects the duration of
transients to which this search is sensitive. 10\,s is a
characteristic timescale for XRF events, although some events of much longer duration have also been observed.

A 3-dimensional sliding {\em celldetect} algorithm is used to search
 for volumes (corresponding to 1 PSF spatially and one transient
 duration temporally) which have a significantly elevated count rate
 compared to their local background. The local background is estimated
 from the count in the adjacent 3 PSF $\times$ 3 transient duration volume
 surrounding the test volume (that is, a spatio-temporal integration
 which is 26$\times$ as large as the test volume).  Poisson statistics
 are used to determine the significance of the local counts given the
 background.  Bins in which the fractional value of the detector
 exposure map is $<$0.35 (where 1.0 is an exposure time equal to the
 total detector exposure time) are ignored.

The algorithm calculates the local count significances throughout the
binned cube, with detection positions overlapping by 4/11 of the PSF
diameter and by 4/5 of the transient duration.  As the \xmm\ PSF
radius changes only modestly with off-axis angle, a changing PSF size
will not affect our results.

A detection threshold significance is chosen so 0.01 false detections are
expected per observation. Test volumes with a significant local
elevation of counts are noted for later analysis.

\subsection{\label{sec: source characterisation}Source Characterization}

The list of candidate XRF events for each observation (typically
containing $\sim$300 events) is then reduced.  Each observation
contains periods of high background, in which the count rate may be
more than an order of magnitude higher than the average value during
the observation. These periods often have a rise time comparable to
the transient search length (see Figure~\ref{fig: event times}), and
give rise to spurious detections.  We therefore remove sources that
occur within high background periods.  A high background period is
found (and defined) as follows:

\begin{enumerate}
\item Split the detector into a 10$\times$10 grid of rectangular spatial regions. 
\item Find the time-averaged countrate for the entire observation in each region. 
\item For each region, rebin the event list into time periods of one
nominal transient duration (10\second). Any bin with a count rate
$>3\times$ the time-averaged count rate for that region is flagged as
a possible high background bin for that region.
\item To avoid counting a very bright transient as a background flare, only
time bins in which $>$ 4 regions are flagged as possible high background
are flagged as whole-chip background flares.
\item Steps 2--4 are repeated three times, taking the time average only over
periods not flagged as high background. 
\end{enumerate}
If necessary, the remaining sources are flagged as redundant - any
particular transient is likely to be detected more than once due to
the spatial and temporal overlaps in the detection algorithm.
Candidate events close to the edge of chips (defined as sources with
PSFs containing regions with fractional exposure $<$0.35) are removed.

Poisson variations in count rate within a few PSF distance of a bright
X-ray source can also produce false detections because of the
relatively large spatial variation in the countrate inside our
background box.  These bright sources
are found manually by the presence of a great (1000+) excesses of
false detections around a bright source.  All candidate events in a
circular area with a radius of 7 PSFs centered on the bright source
are removed; in practice, this step is most efficiently performed by
setting that portion of the detector exposure map to zero in the
initial {\em celldetect} algorithm.  A final manual inspection of the
remaining sources removes sources near obvious detector anomalies,
such as flickering pixels. 

The candidate events which remain after this filtering we consider
to be confirmed events. 

\subsection{Detection Sensitivity}
To determine our detection sensitivity, we performed the following
simulations.

Five-thousand test transients were initially added to each dataset
separate from the detection runs. In each test transient photon-counts were
distributed following Gaussian spatial and time profiles (90\%
diameter of 1 PSF and 90\% duration of 10\second). The transients
were randomly distributed in a 30' diameter circle centered on the
telescope axis, carefully avoiding overlapping in the space-time
volume.

The numbers of test transients detected by the
{\em celldetect} algorithm at an incoming flux per transient of
7.1\tee{-13} \cgsflux\ (10 total counts, on-axis) and
separately at 1.4\tee{-11} \cgsflux\ (200 total counts,
on-axis) were recorded.  An XRF photon spectral index of 2 was assumed
and fluxes are measured over one detection time bin, 2.34\second.

Off-axis exposure and PSF radius variations were included, as were
the effects of chip edges, hot pixels and other detector anomalies.

The numbers of test transients detected at each flux level for each
observation are given in Table \ref{tab: detection
efficiency}. Observations with low detection sensitivities ($<100$
test transients detected) had sensitivity uncertainties reduced by
repeating the Monte Carlo trials several times.  Two observations
(0114120101 and 0097820101) had very low ($<0.05$\%) detection
sensitivities at the 10-count level; we do not include these
observations in our calculations at that flux level.

\section{Results}

\subsection{Candidate Events}

We found two candidate events -- non-Poissonian excesses of counts over
their local background -- in the \xmm-EPIC-pn observations. The
characteristics of the detections are given in Table~\ref{tab:
events} and a detailed description is given in figures \ref{fig:
0055_1} \& \ref{fig:  01253_1}.  Figure \ref{fig: event times} illustrates
that neither of our events were due to background flares. 

To calculate the total number of counts detected over both events, we
correct for the differing exposures of the source and background
regions. This small (O{[}5\% of counts{]}) effect is due to exposure
variations across the telescope.  Correcting for this effect, we find
a total of $15\pm3.8$ counts, including a background of $0.37 \pm
0.16$ counts on the pn detector.

It should  be noted that the candidate events occur within 2 PSFs
of chip edges, and as such may be chip
edge effects.  Additionally, all counts comprising the event in
observation 0125300101 are within one detector time bin (73 ms), which
could also suggest a detector effect.

\subsection{Non-Confirmation of the Astrophysical Nature of the
Confirmed Events, with the MOS Detectors}

The \xmm-EPIC/MOS detectors observe the same area of sky fully
simultaneously with the pn detector.  Therefore, flux from a transient
observed with the pn detector should also be detected by the MOS
chips, albeit at a lower count rate.  

Integrating over both of the confirmed events in the pn detector, a
total of 0 counts were detected by the sum of both MOS detectors at
the same sky locations (1 PSF spatial bin) and times (10\,s time bin), 
in a total estimated local background of 0.55$\ppm$0.16
counts.  Correcting for the fractional exposure difference
between the MOS and PN detectors (but not their different detector
efficiencies) gives a background of 0.52\ppm0.15 counts.

We compare the detected MOS counts with those expected from the pn
counts.  The relative efficiency of the two MOS detectors, for a
source of photon spectral slope of $\alpha$=1.2 is 50\%.  Thus, we
would expect 7.5\ppm1.9 counts in the MOS detectors above background,
instead of the $<2$ counts (90\% confidence) observed.  We therefore place
an upper-limit of $<27$\% of the detected confirmed pn events being
astrophysical in origin -- i.e. neither of our events can be confirmed 
at the 3-$\sigma$ level. 

\subsection{Derived Limits on the Full-Sky XRF Event Rate}

To obtain the upper-limits on the full sky XRF rate, we take our
observed upper-limit to be $<$1 event at the two transient fluxes
described above.  To calculate our total exposure, we use the detector
area (a circle of radius 15'), the total ontime and the detection
fraction from Table~\ref{tab: event rates}.  Since our detection
fractions assumed the spatial distribution to be homogenous across the
detector, these fractions take into account the presence of bright
sources, low-exposure pixels, edge effects and lost detector
columns. The fractions also include sensitivity losses due to ontime
during which background flares occured. The total exposure is
therefore 0.19 sq deg $\times\sum_i (T_{\rm ontime} \times D_{\rm
fraction})$, where $i$ is for each of six observations, $T_{\rm
ontime}$ is the time per observation, and $D_{\rm fraction}$ is the
event detection fraction, which is a function of the number of counts
per event.

At the flux limit corresponding to 10 counts the total exposure is
0.19 sq deg $\times$ 1700\second; at 200 counts, the total exposure is
0.19 sq deg $\times$ 101000\second. The correction factors to
\totalevents\ are 5.5\tee{8} and 8.0\tee{6} respectively.  The
corresponding XRF event rate limits are therefore $<$4.0\tee{9}
\totalevents\ at 10 counts per transient and $<$6.78\tee{7}
\totalevents\ at 200 counts per transient.  This number of counts per
time-bin corresponds to average fluxes of 7.1\tee{-13} \cgsflux\ and
1.4\tee{-11} \cgsflux\, respectively,
 assuming a spectral power-law photon index $\alpha$=2.

Figure \ref{fig: logN logS} shows the relation of our upper limits to
a previous result \cite{vikhlinin98}, as well as to an extrapolation
from the \sax/WFC XRF rate assuming a homogenous source population.
Our limit is less stringent than the previous result by a factor of
$\sim2000$, due in large part to the factor $\sim$13 greater size FOV
of the ROSAT/PSPC detector over the pn detector and the longer
integration time of the previous result.  The limits are also
consistent with the extrapolation of the \sax/WFC XRF rate assuming a
homogenous source population and a power-law photon index of $\alpha$=1.2
 - although it should be noted that we
expected to detect $\sim$ 8 events if the XRFs have a power-law photon
index of $\alpha$=2.

\section{Discussion and Conclusions}

We conduct searches for flashes in the XMM-Newton public archival
data, specifically those observations with total observation time
$>50\,{\rm ks}$ and with galactic latitude $<-15\deg$ or $> 15\deg$.
We use a {\em celldetect} algorithm extended to include three
dimensions (two spatial, one temporal).  Detected sources are
categorized, and non-astronomical sources flagged. Two candidate
events were found in EPIC/pn data, with a total of 15\ppm3.8 counts in
a total estimated background of 0.37\ppm0.16 counts.  The
astrophysical nature of these candidate events is not confirmed by
fully simultaneous observations with the EPIC/MOS detectors, and we
conclude that we have detected no astrophysical XRFs.  We suggest that
the events as observed may be due to detector effects.  

From this we place full-sky, 90\% confidence upper-limits on the XRF
event rate of $<$4.0\tee{9} \totalevents\ (at 7.1\tee{-13} \cgsflux)
and $<$6.8\tee{7}~\totalevents\ (at 1.4\tee{-11} \cgsflux).  The
high-flux limit is above a previous limit obtained with \rosat, by a
factor of $\sim$1500, due to the larger FOV and integration time of
the PSPC data.  The limit at the lower flux remains above that
extrapolated from the \sax/WFC events assuming homgeneity.

The \rosat\ limit remains the most stringent limit on an extrapolation
of the XRF burst rate to lower-fluxes, and implies that the assumption
of homogeneity is violated at a flux $>$2\tee{-11} \cgsflux. To obtain
a limit similar in magnitude to the \rosat\ limit, using XMM/pn, a
total integration time of $\sim$2\tee{8} seconds is required, which
will likely not be obtained with this instrument due to finite
instrument lifetime. It is unclear whether  the XRFs observed with
\sax/WFC are consistent or inconsistent with homogeneity; and so it
may be that a break in the number-fluence distribution, as observed in
GRBs, takes place at a flux at or above that probed by the \sax/WFC
observations.

%%%%%%%%%%%%%%%%BIBLIOGRAPHY%%%%%%%%%%%%%%%%%%%%%%%%%%%%%%%%%%%%%%%%%%%%%%%%%%%%%%%%%%%%%%

\newpage
%\bibliographystyle{astro} \bibliography{complete}

%%%%%%%%%%%%%%%%FIGURES%%%%%%%%%%%%%%%%%%%%%%%%%%%%%%%%%%%%%%%%%%%%%%%%%%%%%%%%%%%%%%
\clearpage

%% Figure 1
%%\pagestyle{empty}
%%\begin{figure}[htb]
%%\caption{\label{fig: event posns}Detected transient positions on the detector
%%chip. The chip layout shown is from the raw count positions of observation
%%097820101.}
%%\end{figure}

%% Figure 1
\pagestyle{empty}
\begin{figure}[htb]
\caption{\label{fig: event times}Detected transient times. Total detector
light curves are shown, binned into the detection time bins of 2.34
seconds. Detected transient times are shown by arrows. The large positive
excursions are due to total-detector increases in the background
countrate.  Observation 0114120101 has a generally elevated count rate
because of a very bright source in the center of the frame. The
periods with reduced count numbers in 0114120101 are due to CCD dead
time.}
\end{figure}

%% Figure 2
\pagestyle{empty}
\begin{figure}[htb]
\caption{\label{fig: logN logS}Cumulative histogram of numbers of detected GRBs and XRFs as a 
function of fluence. We show our upper limits with bars and arrows. The BATSE triggered GRBs 
are normalised to 666 flashes/year/full sky at BATSE's minimum detection fluence \protect\cite{pac99}.
 The BeppoSAX and BATSE XRFs normalisation is approximated by the same factor, after taking 
the different  exposure times of the datasets onto account. An extrapolation of a $-3/2$ 
power-law is shown from the XRFs detected with BeppoSAX and BATSE. XMM (a) is our 90\% upper 
limit at 10 counts per XRF, XMM (b) is our 90\% upper limit at 200 counts per XRF. Both limits 
are derived with an energy pass-band of 1.4-15 keV. The figure also shows the 90\% ROSAT upper 
limit with an energy pass-band of 0.5-2 keV \protect\cite{vikhlinin98}.}
\end{figure}

% Figure 1
%%\clearpage
%%\pagestyle{empty}
%%\begin{figure}[htb]
%%\PSbox{event_raw_posns.ps hoffset=-80 voffset=-80}{14.7cm}{21.5cm}
%%\FigNum{\ref{fig: event posns}}
%%\end{figure}

% Figure 1
\clearpage
\pagestyle{empty}
\begin{figure}[htb]
\PSbox{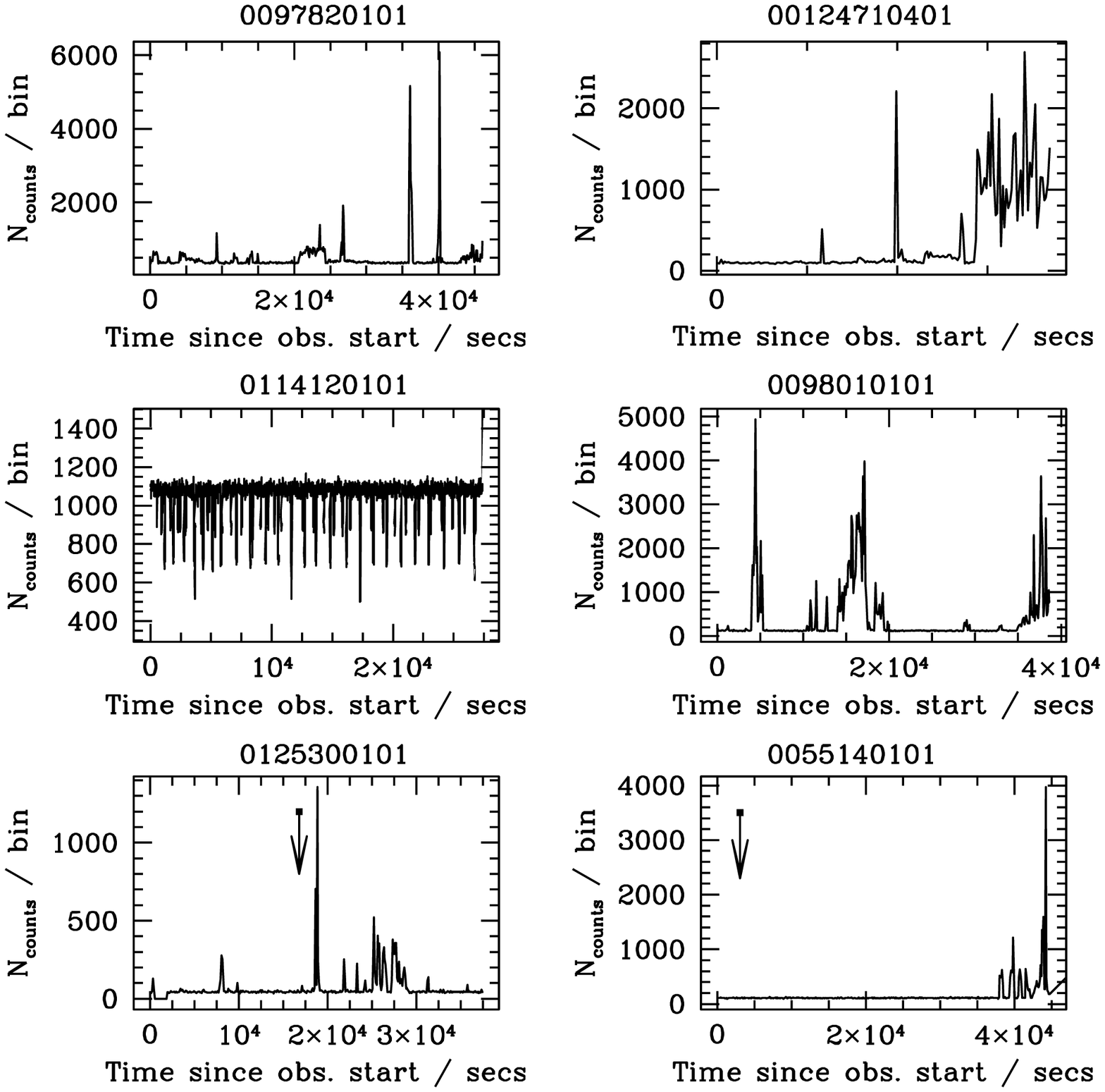 hoffset=-80 voffset=-80}{14.7cm}{21.5cm}
\FigNum{\ref{fig: event times}}
\end{figure}

% Figure 2
\clearpage
\pagestyle{empty}
\begin{figure}[htb]
\PSbox{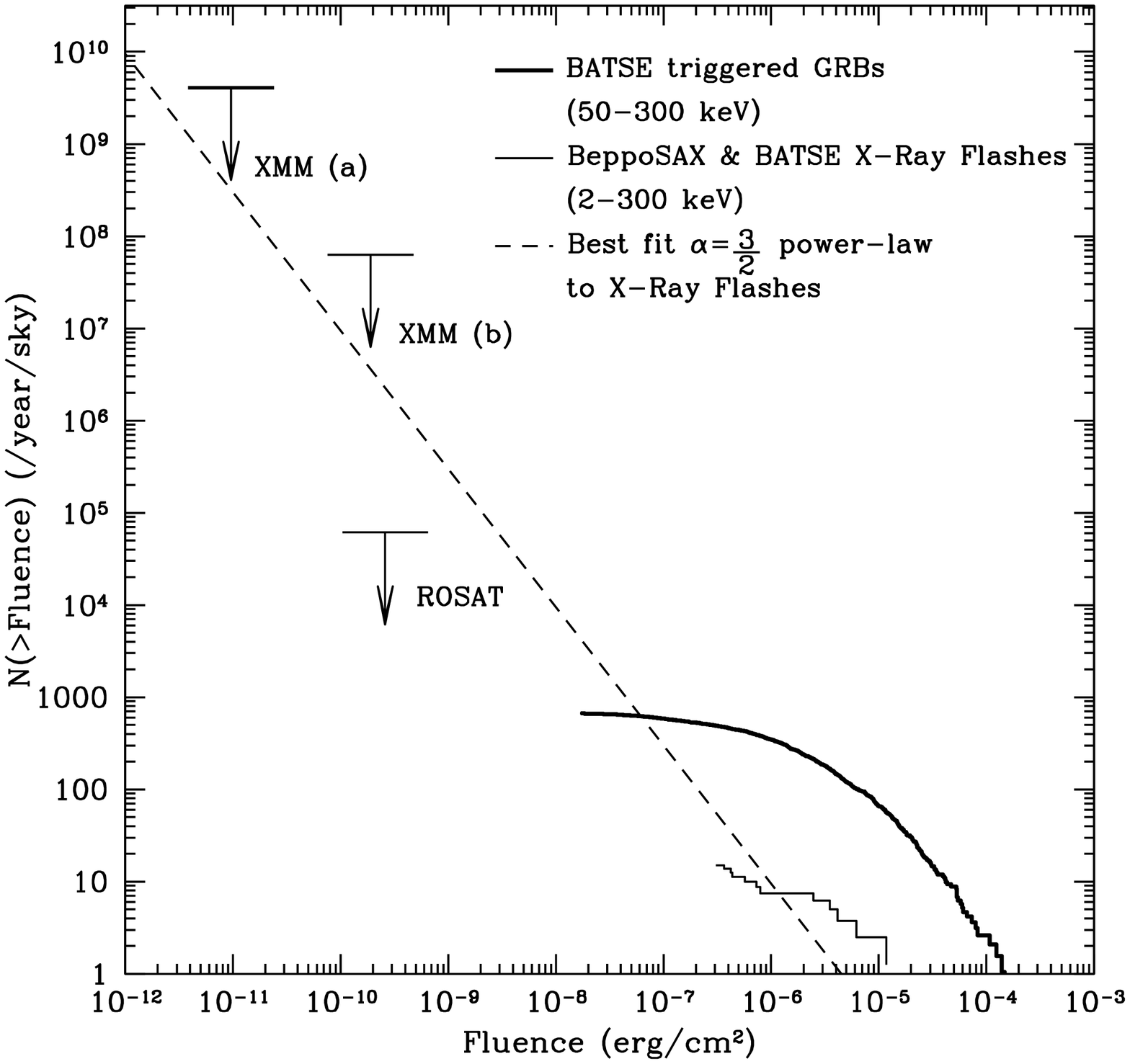 hoffset=-80 voffset=-80}{14.7cm}{21.5cm}
\FigNum{\ref{fig: logN logS}}
\end{figure}

\clearpage

%%%%%%%%%%%%%%%%TABLES%%%%%%%%%%%%%%%%%%%%%%%%%%%%%%%%%%%%%%%%%%%%%%%%%%%%%%%%%%%%%%

%%% Table 1
\begin{table*}
{\centering \begin{tabular}{|c|c|c|c|c|c|c|}
\hline 
Detector&
\multicolumn{2}{c|}{ Chandra ACIS}&
\multicolumn{2}{c|}{XMM EPIC PN }&
\multicolumn{2}{c|}{XMM EPIC MOS }\\
\hline
Flash photon index&
 \( \alpha =1.2 \)&
 \( \alpha =2.0 \)&
 \( \alpha =1.2 \)&
 \( \alpha =2.0 \)&
 \( \alpha =1.2 \)&
 \( \alpha =2.0 \)\\
\hline
Events/year/sky&
 6.3\tee{7}&
 1.5\tee{7}&
 1.8\tee{7}&
 4.8\tee{8}&
 2.2\tee{6}&
 3.0\tee{7}\\
\hline
Expected events&
 0.077&
 1.8&
 0.30&
 8.2&
 0.037&
0.52  \\
\hline
\end{tabular}\par}
\caption{\label{tab: feasibility}Extrapolated event numbers based on flash
detection numbers from WFC BeppoSAX, assuming a minimum of 1 count
per second required for detection. Flash spectra are calculated using
a power law model, $I(E)\propto E^{-\alpha }$;
the dataset was chosen as described in section \ref{sec: data set}. The 
expected event numbers are given for all datasets taken together. The 
$\alpha=1.2$ model may imply a spectral cut-off in XRFs, so avoiding
fluence in the high energy band.
}
\end{table*}

%%%%%%%%%%% Table 2 
\begin{table*}
{\centering {\scriptsize \begin{tabular}{|c|c|p{2cm}|p{1.9cm}|p{2.2cm}|c|}
\hline 
Observation \#&
Total ontime (s)&
Total good time (s)&
\centering Detections at 10 counts&
\centering Detections at 200 counts&
\# trials\\
\hline
\hline 
0055140101&
45257.2&
41266.6&
0.51\ppm0.05\%&
61.8\ppm1.1\%&
8,332,408\\
\hline 
0097820101 {*}&
47721.2&
37620.9&
0.040\ppm0.009\%&
38.4\ppm0.9\%&
7,231,281\\
\hline 
0098010101 {*}&
42121.1&
27760.7&
0.22\ppm0.03\%&
32.2\ppm0.8\%&
5,144,553\\
\hline 
0114120101 {*}&
30453.4&
30453.4&
0.0\%&
46.8\ppm1.0\%&
5,982,074\\
\hline
0124710401&
19939.2&
12505.8&
0.98\ppm0.14\%&
41.1\ppm0.9\%&
1,211,285\\
\hline 
0125300101&
39599.4&
28526.8&
2.9\ppm0.2\%&
47.8\ppm0.98\%&
4,983,749\\
\hline
\end{tabular}}\scriptsize \par}
\caption{\label{tab: detection efficiency}Test transient detection
efficiency.  Good time is the total time of sections of the observation outside
background flares, as defined in section \ref{sec: source
characterisation}. Transients are only detected in the good time -
note that test transients are placed thoughout the ontime.
10 counts corresponds to an on-axis flux of 7.1\tee{-13}
\cgsflux\, 200 counts to 1.4\tee{-11} \cgsflux .  Test photon-counts are
placed in a Gaussian distribution, with 90\% within 5 seconds and 1
PSF of the test transient location. The number of trials given is the number of
individual tests for significance made on the observation; the significance of 
a particular foreground count excess is multiplied by the number 
of trials to obtain its probability. The same number of trials is made
in both the test transient detection and the observed transient
detection and depends on both observation duration and the details of
the observation's exposure map. The number has been corrected to
include only trials in the good time. Observations marked with {*}
have central bright sources removed. 0097820101 has an additional
off-axis bright source removed.}
\end{table*}

%%% Table 3

\begin{table*}
%\begin{sideways}
{\tiny \begin{tabular}{|p{1.3cm}|p{1.0cm}|p{0.6cm}|p{1.5cm}|p{1.5cm}|p{1.05cm}|p{1.1cm}|p{1.1cm}|p{1.1cm}|p{1.1cm}|p{1.0cm}|c|}
\hline 
{Observation \#}&
 {Target}&
 {Event \#}&
 {Expected counts PN/MOS}&
 {Detected counts PN/MOS}&
 {$\left\langle E_{{count}}\right\rangle$ / ${\rm keV}$}&
 {$F_{{peak}}$ / ${\rm ergs / cm^{2}}$ / ${\rm sec}$}&
 {$Fluence$ / ${\rm ergs/cm^{2}}$}&
 {$\left\langle RA\right\rangle$}&
 {$\left\langle DEC\right\rangle$}&
 {$\sigma (pos^{n})$ (statistical uncertainty) /
{}``}&
 {$T_{0}$ / sec}\\
\hline
{0055140101}&
 { LP 944-20}&
 {1}&
 {$0.15\pm 0.11$/ $0.20\pm 0.10$}&
 {$7.0\pm 2.6$/ $0$}&
 {4.73}&
 {8.11\tee{-13}}&
 {6.65\tee{-12}}&
 {03:40:17.5}&
 {-35:20:15.8}&
 {16.8}&
 {95271781}\\
\hline
{0097820101}&
 {A 1795}&
 {-}&
 {-}&
 {-}&
 {-}&
 {-}&
 {-}&
 {-}&
 {-}&
 {-}&
 {-}\\
\hline
{0098010101}&
 {A 1835}&
 {-}&
 {-}&
 {-}&
 {-}&
 {-}&
 {-}&
 {-}&
 {-}&
 {-}&
 {-}\\
\hline
{0114120101}&
 {M87}&
 {-}&
 {-}&
 {-}&
 {-}&
 {-}&
 {-}&
 {-}&
 {-}&
 {-}&
 {-}\\
\hline
{0124710401}&
 {Coma 4}&
 {-}&
 {-}&
 {-}&
 {-}&
 {-}&
 {-}&
 {-}&
 {-}&
 {-}&
 {-}\\
\hline
{0125300101}&
 {J104433.04-012502.2}&
 {1}&
 {$0.22\pm 0.11$ / $0.35\pm 0.12$}&
 {$8.0\pm 2.8$ / $0$}&
 {6.92}&
 {3.2\tee{-12}}&
 {7.6\tee{-12}}&
 {10:44:4.0}&
 {-1:22:20.1}&
 {12.3}&
 {75915279} \\
\hline
\end{tabular}
}
%\end{sideways}

\caption{\label{tab: events}Detected events. MOS counts have been corrected
for the fractional exposure difference between the PN and MOS detectors,
but not for the detectors' relative efficiencies. Peak flux is calculated
over one detection time bin (2.34 s). Flux and fluence assume a photon
index $\alpha =2$ and a passband of 0.4-15keV.
The spectral index is the dominant uncertainty in the flux and fluence,
with energy conversion factors ranging from 4.3\tee{-12} \erg\percm\persec\mbox{${\rm{count}^{-1}}$}
for $\alpha = 1.2$ to 1.2\tee{-13}\erg\percm\persec\mbox{${count}^{-1}$}
for $\alpha=3$. Right ascension and declination are J2000.0, FK5.}
\end{table*}

%%% Table 4 

\begin{table*}
{\tiny\begin{tabular}{|c|c|p{1.5cm}|p{1.5cm}|p{1.3cm}|p{2.4cm}|p{1.5cm}|p{1.5cm}|}
\hline 
{Spectrum \( \alpha  \)}&
 {${\rm R\equiv \frac{c(MOS1\&2)}{c(PN)}}$}&
 {PN counts / BG}&
 {MOS counts / BG}&
 {MOS: predicted counts}&
 {}{90\% upper limit of astrophysical events}&
{10 count event rate / sky / year}&
{200 count event rate / sky / year}\\
\hline
{-1}&
 {0.48}&
{\( 15.0\pm 3.8 \)} {/ \( 0.37\pm 0.16 \)}&
 {\( 0.0\pm 0.0 \)/ \( 0.52\pm 0.15 \)}&
 {\( 7.5\pm 1.8 \)}&
{25\%}&
{\( 6.2x10^{9} \)}&
{\( 1.0x10^{8} \)}{\par}

\\
\hline
{0}&
 {0.57}&
 {\( \cdots  \)}&
 {\( \cdots  \)}&
 {\( 8.8\pm 2.2 \)}&
{21\%}&
{\( 6.2x10^{9} \)}&
{\( 1.0x10^{8} \)}{\par}

\\
\hline
{1}&
 {0.54}&
 {\( \cdots  \)}&
 {\( \cdots  \)}&
 {\( 8.4\pm 2.1 \)}&
{22\%}&
{\( 6.2x10^{9} \)}&
{\( 1.0x10^{8} \)}\\
\hline
{2}&
 {0.32}&
 {\( \cdots  \)}&
 {\( \cdots  \)}&
 {\( 5.2\pm 1.2 \)}&
{37\%}&
{\( 6.2x10^{9} \)}&
{\( 1.0x10^{8} \)}\\
\hline
{3}&
 {0.19}&
 {\( \cdots  \)}&
 {\( \cdots  \)}&
 {\( 3.3\pm 0.7 \)}&
{63\%}&
{\( 6.2x10^{9} \)}&
{\( 1.0x10^{8} \)}\\
\hline
\end{tabular}
}
{\par{}}{\par}

\caption{\label{tab: event rates}The expected 90\% upper flash rate limit
for a number of assumed flash photon indices. The two event fluences
are those given in the detection sensitivity calculations. The PN
to MOS ratio (R) assumes a passband of 0.4-15keV. As our spectral assumptions
indicate neither of the detected events are astrophysical we set the upper 
limit event number to be 1.}
\end{table*}

\section*{\label{sec:appendix}Appendix}

We give detailed information on each of the detections in
Figures \ref{fig: 0055_1} to \ref{fig: 01253_1}. In each figure, we
show: (top left) the low time resolution (2.34 sec/bin) light curve
within 1 PSF of the candidate event position (the central bin, plus
the adjacent bins); (top middle) the high time resolution (0.076 sec)
lightcurve; (top right) the lightcurve of the pn detector; 
(bottom left) the position in the pn detector where
the event was found; (bottom middle) the location of the individual
counts events within the detect cell; and (bottom right) the binned
energy values for the individual count events within the detect cell.

\clearpage
%% Figure A1
\pagestyle{empty}
\begin{figure}[htb]
\caption{\label{fig: 0055_1}0055140101 Event 1.}
\end{figure}

%% Figure A2
\pagestyle{empty}
\begin{figure}[htb]
\caption{\label{fig: 01253_1}0125300101 Event 1.}
\end{figure}

%% Figure A1
\clearpage
\pagestyle{empty}
\begin{figure}[htb]
\PSbox{0055_4.ps hoffset=-80 voffset=-80}{14.7cm}{21.5cm}
\FigNum{\ref{fig: 0055_1}}
\end{figure}

%% Figure A2
\clearpage
\pagestyle{empty}
\begin{figure}[htb]
\PSbox{01253_2.ps hoffset=-80 voffset=-80}{14.7cm}{21.5cm}
\FigNum{\ref{fig:  01253_1}}
\end{figure}

\end{document}